\documentclass[twocolumn,aps,prl,preprintnumbers,bibnotes11pt,superscriptaddress,amsmath,amssymb]{revtex4-1}
\usepackage{graphicx,subfigure}
\usepackage{float}
\usepackage{color}
\usepackage{epstopdf}
\usepackage{amsmath}
\usepackage{amsthm}
\usepackage{amssymb}
\usepackage{amsfonts}
\usepackage{graphicx}
\usepackage{txfonts}
\usepackage{ulem}
\usepackage{mathtools}
\usepackage{bm}

\usepackage{dcolumn}

\usepackage[colorlinks=true,urlcolor=blue,citecolor=blue,linkcolor=red]{hyperref}
\usepackage{booktabs}
\usepackage{braket}
\usepackage[dvipsnames]{xcolor}
\usepackage{ulem}

\usepackage{amsmath}
\usepackage{graphicx}
\usepackage{ulem}
\usepackage{dcolumn}
\usepackage{epsfig}
\usepackage{bm}
\usepackage{array}
\usepackage{hyperref}
\hypersetup{
colorlinks=true,
citecolor=cyan,
linkcolor=blue,
urlcolor=black,
pdfmenubar=true
}

\usepackage{hyperref}
\usepackage{graphicx}
\usepackage{amsmath}
\usepackage{amsfonts}
\usepackage{amssymb}
\usepackage{xcolor}
\usepackage{bbm}
\usepackage{calrsfs}
\usepackage{dutchcal}
\usepackage{amsthm}

\newcommand{\be}{\begin{equation}}
\newcommand{\ee}{\end{equation}}
\newcommand{\beq}{\begin{eqnarray}}
\newcommand{\eeq}{\end{eqnarray}}

\begin{document}
\title{Delta-kick Squeezing}

\author{Robin Corgier}
\affiliation{QSTAR, INO-CNR and LENS, Largo Enrico Fermi 2, 50125 Firenze, Italy.}

\author{Naceur Gaaloul}
\affiliation{Institut f\"ur Quantenoptik, Leibniz Universit\"at Hannover, Welfengarten 1, 30167 Hannover, Germany.}

\author{Augusto Smerzi}
\affiliation{QSTAR, INO-CNR and LENS, Largo Enrico Fermi 2, 50125 Firenze, Italy.}

\author{Luca Pezz$\grave{\text{e}}$}
\affiliation{QSTAR, INO-CNR and LENS, Largo Enrico Fermi 2, 50125 Firenze, Italy.}

\begin{abstract}
We explore the possibility to overcome the
standard quantum limit (SQL) in a free-fall atom interferometer using a Bose-Einstein condensate (BEC) in either of the two relevant cases of Bragg or Raman scattering light pulses.
The generation of entanglement in the BEC is dramatically enhanced by amplifying the atom-atom interactions via the rapid action of an external trap focusing the matter-waves to significantly increase the atomic densities during a preparation stage -- a technique we refer to as delta-kick squeezing (DKS).
The action of a second DKS operation at the end of the interferometry sequence allows to implement a non-linear readout scheme making the sub-SQL sensitivity highly robust against imperfect atom counting detection.
We predict more than 30\,dB of sensitivity gain beyond the SQL for the variance, assuming realistic parameters and $10^6$ atoms. 
\end{abstract}

\maketitle

Free-fall atom interferometers~\cite{Berman1997,Cronin2009,Tino2014} are extraordinarily sensitive to external forces and find key applications as gravimeters, gradiometers and gyroscopes in applied physics as well as in fundamental science~\cite{Bongs2019, arxiv_Geiger_2020, Safronova2018}. 
State-of-the-art devices use $N$ uncorrelated atoms and their phase estimation uncertainty is lower bounded by the the standard quantum limit (SQL), $\Delta \theta_{\rm SQL} = 1/\sqrt{N}$.
Since $N$ is generally constrained by the experimental apparatus or by the onset of unwanted systematic effects due to the high density, the possibility to overcome the SQL by engineering specific quantum correlations~\cite{PRL_Pezze_2009} between the atoms is attracting increasing interest~\cite{MRP_Pezze_2018}.

While proof-of-principle entanglement-enhanced atom interferometry~\cite{MRP_Pezze_2018, Nature_Riedel_2010, Nature_Gross_2010, Science_Lucke_2011,
BohnetNATPHOT2014, Nature_Hosten_2016} have been largely investigated both theoretically and experimentally in the context of atomic clocks~\cite{Louchet-ChauvetNJP2010, LerouxPRL2010, PRL_Kruse_2016,PedrozoNATURE2020,AndrePRL2004, SchulteNATCOMM2020, PRL_Pezze_2020} and magnetometry~\cite{PRA_Petersen_2005,PRL_Wasilewski_2010,PRX_Ruster_2017,OckeloenPRL2013, MuesselPRA2014, SewellPRL2012}, free-fall atom interferometers have received surprisingly less attention~\cite{PRL_Haine_2013, PRL_Hamilton_2015,PRL_Salvi_2018, QST_shankar_2019,PRL_Szigeti_2020,arxiv_Anders_2020}. 
The main reason is that these measurement devices have stringent practical requirements: in particular, the generation of atomic entanglement must be compatible with the splitting of the atomic wave-packets in momentum modes. Bose-Einstein condensates (BECs) have been pinpointed as optimal candidates for the realization of entanglement-enhanced free-fall atom interferometers~\cite{PRL_Szigeti_2020}. Indeed,
the narrow momentum dispersion guarantees ideal splitting~\cite{NJP_Szigeti_2012} and entanglement can be generated via particle-particle interactions~\cite{MRP_Pezze_2018, Nature_Sorensen_2001, Nature_Riedel_2010,Nature_Gross_2010, PRA_Kitagawa_1993, PRL_Szigeti_2020}.
However, since the interaction vanishes due to free-fall expansion after a short transient time prior to the interferometer operations, current theoretical studies predict only a modest sub-SQL sensitivity gain~\cite{PRL_Szigeti_2020}. 

In this paper, we overcome these limitations by proposing a novel method to enhance the generation of entanglement in free-fall atom-interferometers using BECs. The key idea consists in focusing the matter-waves through the rapid application of an external trapping potential in analogy to optics, where the trap plays the role of a converging lens. 
Going through the focal point increases the matter-wave density and thus the effective strength of the particle-particle interactions preparing the atoms in a highly-entangled spin-squeezed state.
Considering previous works on delta-kick collimation~\cite{OptLettChu1986,PRLAmmann1997,Muntinga2013,PRLKovachy2015,NJP_Corgier_2020}, we designate our technique {\it delta-kick squeezing} (DKS). 
The method is explored for Raman and Bragg scattering and is made fully compatible with the requirement of linear atom interferometer operations. The DKS technique leads to a substantial phase sensitivity gain beyond the SQL, e.g. more than 30\,dB in a realistic experimental configuration with $10^6$ atoms.
For Bragg diffraction, in particular, using a second DKS pulse at the end of the interferometer sequence allows the realization of a nonlinear readout protocol~\cite{LeibfriedNATURE2005, PRL_Davis_2016, PRL_Frowis_2016, PRA_Macri_2016, Science_Hosten_2016, NolanPRL2017, AndersPRA2018, Schulte_2020}. In this case, the twisting dynamics generating the spin-squeezed state is inverted before the final measurement of atom numbers in the two interferometer output ports. 
This operation makes the interferometer exceptionally robust against detection noise.

\begin{figure*}[t!]
\includegraphics[width=\textwidth]{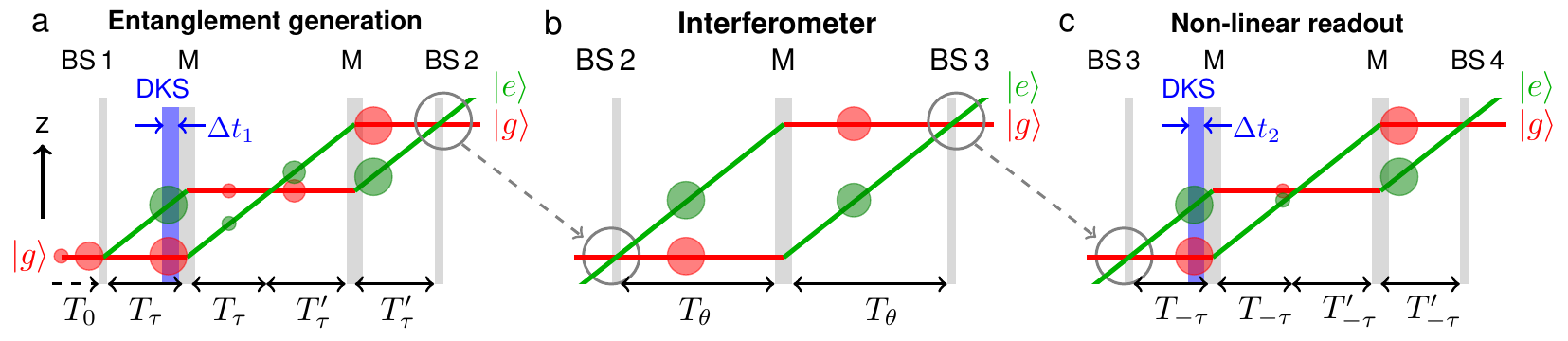}
\caption{Complete operation of a DKS-enhanced free-fall atom interferometer. State preparation (a) consists of i) a first expansion of the BEC after release from the trap, $T_0$; ii) a beam-splitter (BS1); iii) a DKS of duration $\Delta t_1$ focusing the matter-waves; and iv)  two mirror pulses (M). The inteferometer sequence (b) starts with the beam-splitter pulse (BS2) followed by a mirror (M) and ending with a recombining beam-splitter (BS3). The pulses are equally separated in time by $T_{\theta}$. In the case of a linear detection scheme, the phase is evaluated by counting the number of atoms at the two output ports. In a non-linear readout case (c), one applies operations analogue to the ones of the state preparation. Right before the first mirror pulse, a second DKS of duration $\Delta t_2$ is applied and generates an "un-twisting" dynamics. The phase is finally evaluated by counting the number of atoms at the two output ports.}
\label{fig_1}
\end{figure*}

{\it Entanglement enhancement by the DKS state preparation protocol.}
The preparation step illustrated in Fig.~\ref{fig_1}(a) starts with a BEC suddenly released from an external trap. 
A short free expansion time $T_0$
dilutes the BEC and guarantees, by applying a first beam splitter pulse (BS1), the preparation of the quantum superposition $\ket{\psi_0} = (\ket{g} + \ket{e})^{\otimes N}/2^{N/2}$, $N$ being the number of atoms, and $|g\rangle$ and $|e\rangle$ two momentum states \cite{Raman_vs_Bragg}.
Entanglement is then generated via particle-particle interactions in the BEC, such that $\ket{\psi_0}$ evolves according to the one-axis twisting dynamics  $|\psi_{sq}(t)\rangle = e^{-i\tau(t)\hat{S}_z^2}|\psi_{0}\rangle$
\,\cite{PRA_Kitagawa_1993,Nature_Sorensen_2001, Approximation2}, where $\tau(t)=\int_0^{t} \chi(t')\, dt'$.
Notably, the time dependent non-linear coefficient $\chi(t)$ is given by~\cite{supp}
\begin{subequations}
\label{eq_chi}
\begin{align}
\label{eq_chi_R}
\chi^{R}(t)&=\chi_{S}(t)-\chi_{C}(t),\\
\label{eq_chi_B}
\chi^{B}(t)&=\chi_{S}(t)-2\chi_{C}(t),
\end{align}
\end{subequations}
for the Raman (R) or the Bragg (B) scattering, respectively.
Here, $\chi_{S}(t)=g \int_{-\infty}^{\infty}\,d\mathbf{r}\,\left|\Phi_{g (e)}(\mathbf{r},t)\right|^4/\hbar$ and $\chi_{C}(t)=g \int_{-\infty}^{\infty}\,d\mathbf{r}\,\left|\Phi_{g}(\mathbf{r},t)\right|^2 \left|\Phi_{e}(\mathbf{r},t)\right|^2/\hbar$ denote the self-phase and cross-phase modulation terms, respectively, where, $\Phi_{g (e)}(\mathbf{r},t)$ carries the spatial evolution of the state $|g\rangle$ ($|e\rangle$). For simplicity, in the following we assume
the same intra- and inter-species scattering coefficient
$g= g_{11}= g_{12}=g_{22}>0$ \cite{Approximation1}.
The factor 2 in front of the cross-phase modulation terms in Eq.~(\ref{eq_chi_B})~\cite{PRA_Guan_2020, arxiv_Corgier_2020} is due to the interference of the two modes (see \cite{supp}) and is rich of consequences. In particular, when the wave functions overlap $\chi_{S}$ and $\chi_{C}$ are equal and thus $\chi^R \approx 0$ in the Raman case~\cite{Nature_Riedel_2010, Nature_Gross_2010}. 
In contrast, during overlap,
$\chi^{B} \approx -\chi_{S} \neq 0$, making the one-axis-twisting evolution active in the Bragg case. 
Furthermore, $\chi^{B}$ can assume either positive or negative values (see below), while $\chi^R\geq 0$.

\begin{figure}[t!]
\includegraphics[width=0.90\columnwidth]{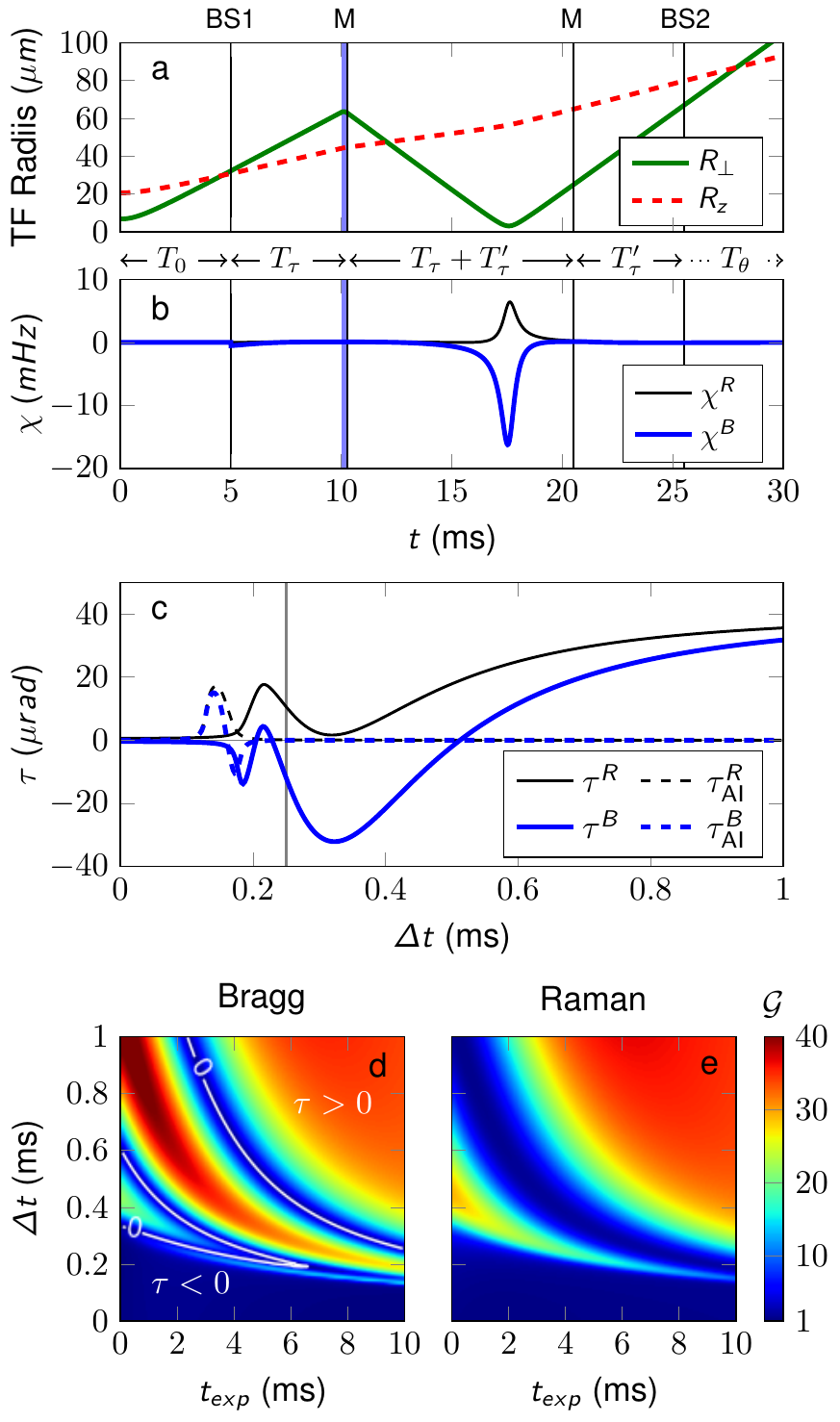}
\caption{DKS engineering. (a) Thomas-Fermi Radii along the z ($R_z$) and transverse directions ($R_\perp$) as a function of time during state preparation (see~\cite{Parameters} for the parameters used). The different laser pulses are highlight by the vertical black lines with $T_{\tau}=T'_{\tau}=5.25$ ms and the DKS time is fixed to  $\Delta t=0.25$\,ms (vertical blue lines). (b) Corresponding non-linear coefficient $\chi(t)$ for Raman and Bragg scattering.
(c) Effective non-linear coefficient, $\tau$, as a function of the DKS duration, $\Delta t$, during the state preparation (solid lines) and during the interferometer sequence (dashed). The vertical line denotes the case shown in panels (a) and (b). Panels (d) and (e) show the sensitivity gain (color scale) for the Bragg (d) and Raman (e) configurations as a function of the  pre-expansion time and DKS duration. The white lines denotes $\tau^B=0$ and distinguished the regions $\tau^B>0$ from $\tau^B<0$. Here\,\,$\mathcal{G}=40$\,\,corresponds\,\,to\,\,a\,\,variance\,\,of\,\,32\,dB\,\,below\,\,the\,\,SQL.}
\label{fig_2}
\end{figure}

The atomic interactions, proportional to the density of the freely-expanding cloud, are vanishing a few milliseconds after release in a free-fall interferometer and therefore prohibit the generation of highly entangled states \,\cite{PRL_Szigeti_2020}.
This problem is overcome here by switching on, after a pre-expansion time $t_{exp}$, an external harmonic trap for a  time $\Delta t$, to induce a size focusing of the atomic cloud (Fig.~\ref{fig_2}a)~\cite{Delta_kick_analogy}. 
This re-focusing increases the density of the cloud and thus the effective interaction coefficient $\tau$.
Right after this delta-kick pulse, at time  $T_{\tau}=t_{exp}+\Delta t_1$, a first mirror pulse (M) is applied. 
Since the DKS also imprints a phase equivalent to a classical center-of-mass motion, the detuning of the mirror pulse has to be adjusted to absorb the change of trajectories~\cite{PRL_Roura_2017,PRL_DAmico_2017,PRL_Overstreet_2018}. 
Finally, the state preparation stage ends after an additional time $2T'_{\tau}$ (notice that a second mirror pulse is applied after a time $T'_{\tau}$) necessary to dilute the cloud density and to guarantee that the interferometer sequence is implemented with non-interacting atoms.
The time $T'_{\tau}$ can be tuned depending on the specific DKS parameters.
In Fig.~\ref{fig_2}(a-c) we present a realistic example of state preparation using the DKS~\cite{Parameters}: we show the size of the BEC cloud in Fig.~\ref{fig_2}(a) and the non-linear coefficient $\chi$ as a function of time in Fig.~\ref{fig_2}(b). 
The latter clearly shows how $\chi(t)$ is enhanced by the DKS and is different for Raman and Bragg pulses.
Figure~\ref{fig_2}(c) shows $\tau$ as a function of the duration of the DKS, $\Delta t$. 
In particular, for Bragg scattering, $\tau$ can have either positive or negative values, depending on $\Delta t$.

{\it Atom interferometry with linear detection.}
The Mach-Zehnder interferometer sequence illustrated in Fig.~\ref{fig_1}(b) consists of two beam-splitters (BS2 and BS3) and a mirror pulse (M) equally spaced in time by $T_\theta$.
We define the sensitivity gain over the standard quantum limit, $\mathcal{G}=\Delta \theta_{\rm SNL}/\Delta \theta$ calculated at $\theta=0$~\cite{PRA_Wineland_1994}, where $\theta$ denotes the phase accumulated during the interferometer and $(\Delta \theta)^2 = (\Delta S_z)^2/( d\langle S_z \rangle / d \theta \vert_{\theta=0} )^2$ is obtained by error propagation.
As discussed previously, the linear interferometer condition $\tau_{AI}=\int_{T_i}^{T_i+2T_\theta} \chi(t') dt'= 0$, with $T_i=T_0+2T_\tau+2T'_\tau$, can be realized independently of the DKS parameters, $t_{exp}$ and $\Delta t$, by adjusting the time $T'_\tau$.
In this case, the ideal interferometer sequence can be described by the linear transformation $\hat{U}_{\rm AI}=e^{i\theta\hat{S}_y}$ such that the output state is $\ket{\psi_{out}} = \hat{U}_{\rm AI} \ket{\psi_{sq}}$~\cite{Extra_phase}.
The sensitivity gain hence reads
\beq
\label{eq_Gain}
\mathcal{G}=\dfrac{2\cos(\tau)^{ N-1}}{[4+(N-1)(A-\sqrt{A^2+B^2})]^{1/2}},
\eeq
with $A=1-\cos(2\tau)^{N-2}$ and $B=4\sin(\tau)\cos(\tau)^{N-2}$, upon an opportune rotation~\cite{PRA_Kitagawa_1993} of the squeezed state $|\psi_{sq}(t)\rangle$ at BS2. The maximum value of $\mathcal{G}$ is reached for $\tau_{\rm opt}\approx{1.2\,N^{-1/3}}$ ~\cite{PRL_Pezze_2009, PRA_Kitagawa_1993}.
In Fig.\,\ref{fig_2}(d,e) we plot the gain $\mathcal{G}$ as a function of the DKS parameters $t_{exp}$ and $\Delta t$, for Bragg and Raman scattering, respectively. 
While for $\Delta t=0$, no significant gain can be obtained ($\mathcal{G}\approx1$), the DKS enables the creation of highly-entangled input states in these freely expanding configurations: a large gain is possible for a large parameters range of pre-expansion, $t_{exp}$, and DKS duration, $\Delta t$. 

\begin{figure}[t!]
\includegraphics[width=\columnwidth]{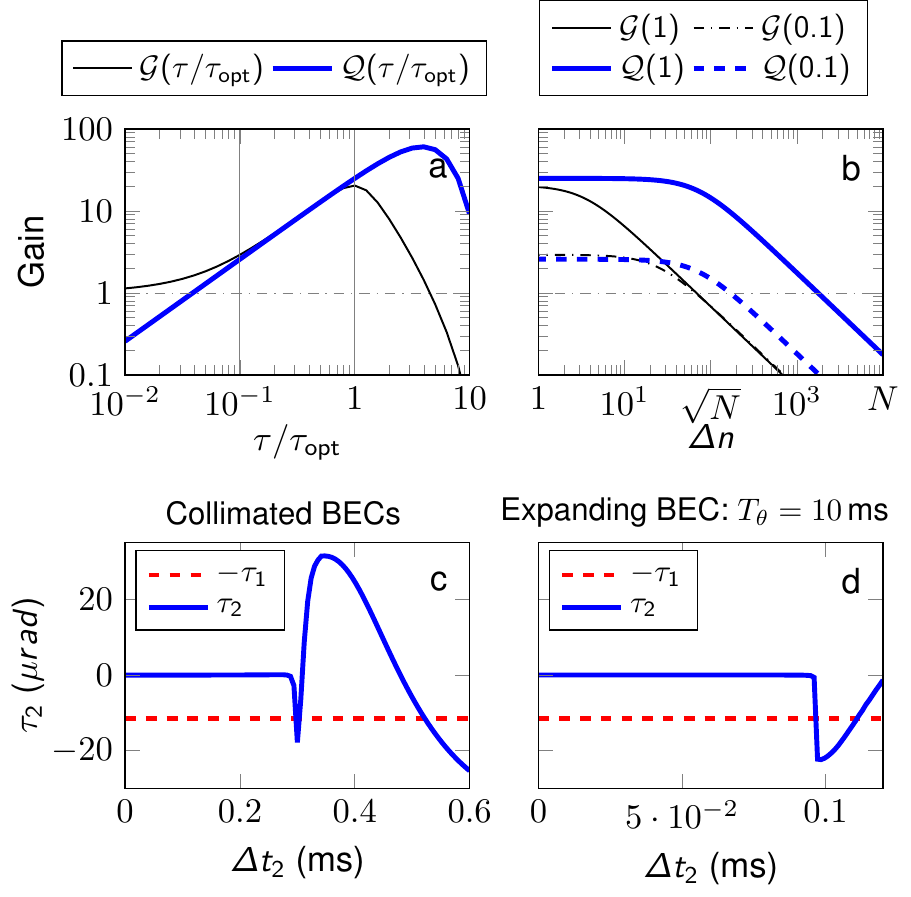}
\caption{Non-linear readout. (a) Sensitivity gain for linear,  $\mathcal{G}(\tau/\tau_{\rm opt})$, and  non-linear, $\mathcal{Q}(\tau/\tau_{\rm opt})$, readout as a function of the effective non-linear coefficient, $\tau=\tau_1=-\tau_2$. The vertical lines denote $0.1 \tau_{\rm opt}$ and $\tau_{\rm opt}$. (b) Sensitivity gain as a function of the detection noise $\Delta n$ for linear and non-linear readout, $\tau=\tau_{\rm opt}$ (solid) and $\tau=0.1\tau_{\rm opt}$ (dashed). The number of atoms is $N=10^4$. Panels (c) and (d) show the non-linear coefficient as a function of the second DKS duration, $\Delta t_2$ (solid blue line). In panel (c) the size of the BECs are assumed to be constant after the BS2 while in panel (d) the BECs continue to expand after BS2. For specific values of $\Delta t_2$, the nonlinear coefficient matches $-\tau_1\approx -0.1\tau_{\rm opt}$~\cite{Parameters2}.}
\label{fig_3}
\end{figure}

{\it Atom interferometry with non-linear readout.}
Using the DSK to tune the sign of the effective interaction for Bragg scattering can be exploited to realize a non-linear readout scheme.
After the interferometer sequence, a second DKS is applied, see Fig.~\ref{fig_1}(c), such that the output state is now given by
\be 
\label{eq_Sq_Phase_UnSq}
|\psi_{\rm out}\rangle =e^{-i \pi/2 \hat{S}_y} e^{-i \tau_2\hat{S}_z^2}\hat{U}_{\rm AI} e^{-i \tau_1 \hat{S}_z^2}|\psi_0\rangle.
\ee
Notice that, for simplicity, the final rotation $e^{-i \pi/2 \hat{S}_y}$ is not included in Fig.~\ref{fig_1}.
For the sake of readability, we distinguish the sensitivity gain of the linear detection, $\mathcal{G}$, to the non-linear readout, $\mathcal{Q}$.
In the case where 
\be \label{condition}
\tau_1=-\tau_2\equiv\tau
\ee
and $\theta=0$, it has already been shown that non-linear readout provides (i) the possibility to reach a higher sensitivity gain, $\rm{max}(\mathcal Q)>\rm{max}(\mathcal G)$, with respect to the linear detection case for sufficiently large values of $\tau$ (see Fig. \ref{fig_3}a)~\cite{PRL_Davis_2016}; and (ii) a phase magnification robust against atom number detection imperfection (see Fig.\,\ref{fig_3}b)\,\cite{PRL_Davis_2016, Science_Hosten_2016}.
The later being one of the most critical limitations in quantum-enhanced atom interferometers \cite{MRP_Pezze_2018}. Throughout this paper, the imperfect detection resolution is modeled by a Gaussian noise of variance $(\Delta n)^2$~\cite{MRP_Pezze_2018,PRL_Szigeti_2020}.

\begin{figure}[t!]\includegraphics[width=\columnwidth]{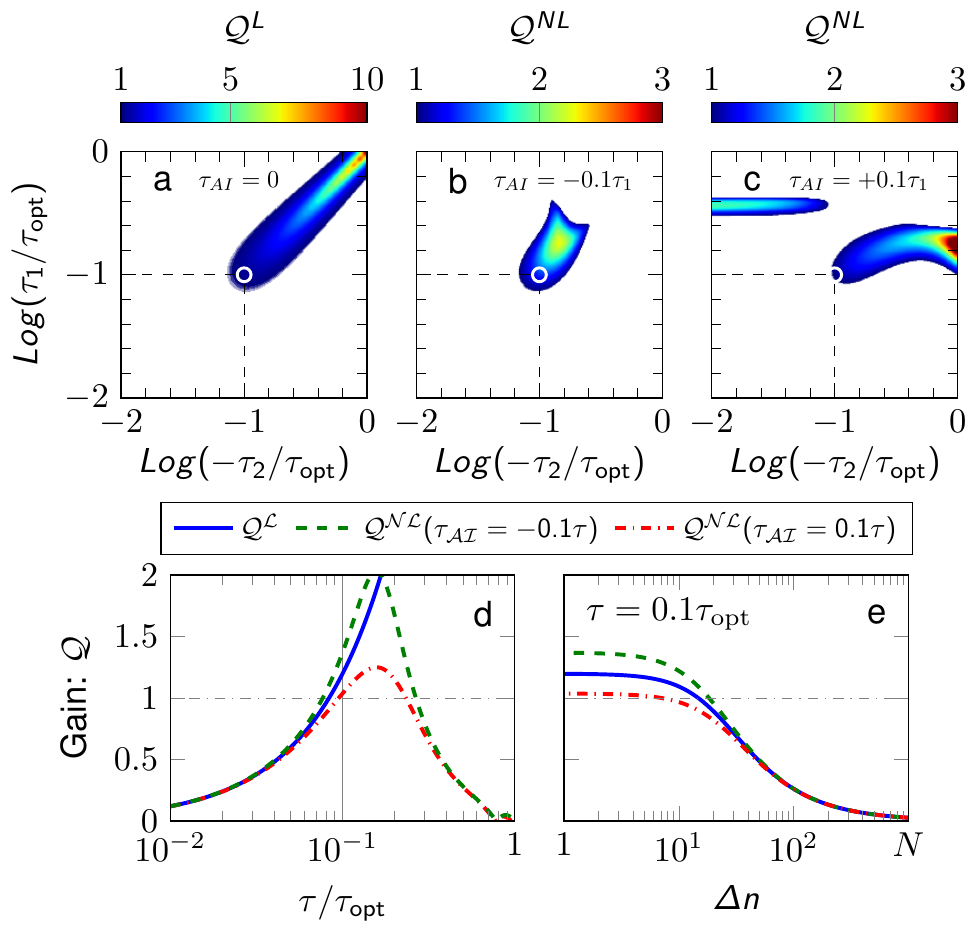}
\caption{Robustness of the non-linear readout. Interferometer sensitvity gain as a function of $\tau_1$ and $\tau_2$ for linear ($\mathcal{Q}^L$, panel a) and non-linear ($\mathcal{Q}^{NL}$) interferometer sequence with $\tau_{AI}=10\%\,\tau_1$ (b) and $\tau_{AI}=-10\%\,\tau_1$ (c). Panel d: Sensitivity gain as a function of $\tau=\tau_1=-\tau_2$. The case $\tau=0.1\tau_{\rm opt}$ corresponds to the circles in panels (a-c) and is highlighted by the vertical line. Panel e: Sensitivity gain as a function of the detection noise parameter $\Delta n$ and for $\tau=0.1\tau_{\rm opt}$. In all panels $N=10^3$. Here\,\,$\mathcal{Q}=10\,(3)$\,\,correspond\,\,to\,\,a\,\,variance\,\,of\,\,20\,\,(9.5)\,\,dB\,\,below\,\,the\,\,SQL.}
\label{fig_4}
\end{figure}

It should be noticed that satisfying the conditions (\ref{eq_Sq_Phase_UnSq}) and (\ref{condition}) is not straightforward in quantum systems as it requires inverting the twisting evolution that generated entanglement in the probe state.
Such a possibility has been predicted for Rydberg atoms~\cite{PRL_Davis_2016, PRA_Macri_2016} and has been experimentally realized for cold atoms in a cavity~\cite{Science_Hosten_2016} and trapped ions~\cite{LeibfriedNATURE2005, GarttnerNATPHYS}. 
Here the condition
(\ref{condition}) can be naturally satisfied
by using Bragg scattering and tuning the DKS parameters.
This is shown in Fig.\,\ref{fig_3}(c-d) where we plot $\tau_2$ as a function of the DKS durations $\Delta t_2$ and compare it to $-\tau_1$. The condition (\ref{condition}) can be satisfied in realistic experimental conditions.
In particular, panel (c) shows the case where the sizes of the BECs are kept constant after BS2 until the second DKS during the non-linear readout. 
This configuration can be engineered through the action of a delta-kick collimation pulse~\cite{OptLettChu1986,PRLAmmann1997,Muntinga2013,PRLKovachy2015,NJP_Corgier_2020} before BS2 when the two clouds are dilute enough.
This manipulation leads to record low-expansion rates as low as 50\,pK~\cite{PRLKovachy2015}  enabling interferometer sequences longer than $T_\theta=1$\,s.
In addition collimated atomic ensembles brings the possibility to perfectly ``un-twist'' the quantum states with a second DKS pulse of some hundreds of microseconds (Fig.\,\ref{fig_3}c) accessible to today's capabilities, while expanding BECs would require shorter DKS pulses of few tens of microseconds (Fig.\,\ref{fig_3}d).
In Fig\,\ref{fig_4}a we give a more complete overview of the nonlinear readout parameters beyond the condition (\ref{condition}). There, we plot
the sensitivity gain $\mathcal{Q}$ as a function of $\tau_1$ and $- \tau_2$. We see that slight unbalanced conditions (namely $\vert \tau_1+\tau_2 \vert \gtrsim 0$) still enable high sensitivity gains.

{\it Impact of residual mean-field interactions.}
In the presence of a small residual interactions after the state preparation, the interferometer transformation is described by $\hat{U}_{\rm AI}=e^{i\theta\hat{S}_y}e^{i\tau_{AI}\hat{S}_y^2}$ in Eq.\,\ref{eq_Sq_Phase_UnSq}.
While the interactions during the interferometer largely degrade the sensitivity gain in a linear detection scheme ($\mathcal{G}^{NL} \leq \mathcal{G}^{L}$~\cite{arxiv_Corgier_2020}, the labels $L$ and $NL$ distinguish between a linear and non-linear interferometer sequence, respectively), this is surprisingly not the case when exploiting the non-linear readout. 
In Fig.~\ref{fig_4}(b) and (c), we plot $\mathcal{Q}^{NL}$ as a function of $\tau_1$ and $-\tau_2$ for $\tau_{AI} = \pm 0.1 \tau_1$ and contrast it to the linear case ($\mathcal{Q}^{L}$, for $\tau_{AI} = 0$) in Fig.~\ref{fig_4}(a). 
We conclude that sub-shot-noise sensitivities are obtained for a small imbalance between $\tau_1$ and $-\tau_2$ when $sign(\tau)\neq sign(\tau_{AI})$ (panel b) or large imbalance when $sign(\tau)=sign(\tau_{AI})$ (panel c).
To emphasize these results, we plot in Fig.~\ref{fig_4}(d) the sensitivity gain for the specific case, $\tau_1=-\tau_2$. 
Here we show that if $sign(\tau) \neq sign(\tau_{AI})$, we can find a parameter range where $\mathcal{Q}^{NL} \geq \mathcal{Q}^L>1$.
On the other hand in this configuration if $sign(\tau)=sign(\tau_{AI})$, $\mathcal{Q}^{NL}<\mathcal{Q}^{L}$.
This result is confirmed analytically in the regime of low interactions where~\cite{supp}
\begin{equation}
\left(\mathcal{Q}^{NL}\right)^2=\frac{1}{4}\left[(4\tau-3\tau_{AI})\dfrac{N}{2}-(2\tau-\tau_{AI})\right]^2.    
\end{equation}
In Fig.~\ref{fig_4}(e) we plot the corresponding robustness to atom number resolution of the non-linear readout for $\tau=0.1\tau_{opt}$ and verify that there is no significant gain difference between the different configurations.  
While the presence of residual interactions and imperfect detection prohibit sub-SQL sensitivity with linear detection schemes, non-linear readout enables the creation of a quantum-enhanced interferometer sequence, in the regime of low interactions, robust against imperfect detection.

{\it Conclusion.}
In this manuscript we have proposed and exploited a phase-space engineering technique to focus matter waves that (i) substantially enhances the amount of entanglement generated in the probe state of a free-fall atom interferometer and (ii) realizes, in the case of Bragg diffraction, a non-linear readout protocol making the interferometer sensitivity extraordinarily robust against detection noise.
This robustness is crucial in atom interferometers with non-classical states to avoid dramatic effects of imperfect atom number detection on sub-SQL sensitivities.
Our predictions have been obtained for realistic experimental parameters and assuming ideal beam-splitter and mirror pulses. 
Different effects may degrade the sensitivity of the interferometer as atomic losses~\cite{PRL_Sinatra_2008,Sinatra_2009}, mode mismatch~\cite{PRA_Poulsen_2002} and shape deformation~\cite{PRA_Burchianti_2020} due to BEC crossing, and residual interaction during splitting pulses~\cite{PRA_Pezze_2006}. A careful analysis of these effects depends on the specific interferometer configuration and is beyond the scope of this work.
The assumption of harmonic traps to realize the DKS is justified for BEC sizes and the proposed spatial separations.
Within these assumptions, a sensitivity gain of more than 30 dB beyond the SQL with $10^6$ atoms is predicted for both Bragg and Raman diffraction. Larger harmonic traps~\cite{PRA_Roy_2016,PRL_Condon_2019} could enable even higher sensitivity gains. 

A straightforward implementation of our technique, can boost the sensitivity of a BEC gravimeter~\cite{Freier2016,Karcher2018} of $10^6$ atoms to that of an ensemble with a hundredfold flux.
Our DKS technique thus promotes BEC ensembles in free-fall atom interferometers to primary quantum-enhanced sensors to explore timely physics quests such as testing General relativity principles~\cite{Battelier2019,Tino_2021b} or atom-interferometric gravitational-wave detection~\cite{Canuel2020}.

\begin{acknowledgments}
RC thanks Franck Pereira Dos Santos and Peter Wolf, NG thanks Christian Schubert and Klemens Hammerer for fruitful discussions. This work is supported by the European Union's Horizon 2020 research and innovation programme - Qombs Project, FET Flagship on Quantum Technologies grant no. 820419. NG acknowledges the support of the CRC 1227 (DQmat) within Project No. A05, the German Space Agency (DLR) for funds provided by the Federal Ministry for Economic Affairs and Energy (BMWi) due to an enactment of the German Bundestag under Grant Nos. 50WM\-1861 (CAL) and 50WM\-2060 (CARIOQA) and the Deutsche Forschungsgemeinschaft (DFG, German Research Foundation) under Germany’s Excellence Strategy – EXC-2123 QuantumFrontiers – 390837967.
\end{acknowledgments}

\widetext
\clearpage
\begin{center}
\textbf{\large Supplemental Materials}
\end{center}
\setcounter{equation}{0}
\setcounter{figure}{0}
\setcounter{table}{0}
\setcounter{page}{1}
\makeatletter
\renewcommand{\theequation}{S\arabic{equation}}
\renewcommand{\thefigure}{S\arabic{figure}}
\renewcommand{\bibnumfmt}[1]{[S#1]}
\renewcommand{\citenumfont}[1]{S#1}

\onecolumngrid
\section{Raman vs Bragg scattering}

\subsection{Two-component single external mode Hamiltonian (Raman)}
Here we consider the two-component non-linear Hamiltonian,
\be
\hat{H}^R_{\rm int}(t)=\dfrac{g_{11}}{2} \int_{-\infty}^{\infty}\,d\mathbf{r}\,\hat{\Psi}_1^\dagger\hat{\Psi}_1^\dagger\hat{\Psi}_1\hat{\Psi}_1+\dfrac{g_{22}}{2} \int_{-\infty}^{\infty}\,d\mathbf{r}\,\hat{\Psi}_2^\dagger\hat{\Psi}_2^\dagger\hat{\Psi}_2\hat{\Psi}_2+g_{12} \int_{-\infty}^{\infty}\,d\mathbf{r}\,\hat{\Psi}_1^\dagger\hat{\Psi}_2^\dagger\hat{\Psi}_1\hat{\Psi}_2,
\ee
where the field operators are $\hat{\Psi}_1=\phi_1(\mathbf{r},t)\, \hat{a}_1$ and $\hat{\Psi}_2=\phi_2(\mathbf{r},t) e^{2ikz}\, \hat{a}_2$. Here $\phi_i(\mathbf{r},t)$ denotes the spatial shape of component $\hat{\Psi}_i$ and $ e^{2ikz}$ is the phase imprinted by the two-photon transition. $\hat{a}_{i}$  ($\hat{a}_{i}^\dagger$) is the bosonic annihilation (creation) operator. It is convenient to introduce the SU(2) pseudo-spin operators of the Lie's algebra,
\begin{eqnarray}
\label{eqA_Sx_Sy_Sz_def}
    \hat{S}_x &=& \left(\hat{a}^\dagger_{1}\hat{a}_{2}+\hat{a}_{2}^\dagger\hat{a}_{1}\right)/2, \nonumber \\
    \hat{S}_y &=& \left(\hat{a}_{1}^\dagger\hat{a}_{2}-\hat{a}_{2}^\dagger\hat{a}_{1}\right)/2i,\\  \nonumber
    \hat{S}_z &=& \left(\hat{a}_{1}^\dagger\hat{a}_{1}-\hat{a}_{2}^\dagger\hat{a}_{2}\right)/2,\\  \nonumber
     \hat{N} &=& \left(\hat{a}_{1}^\dagger\hat{a}_{1}+\hat{a}_{2}^\dagger\hat{a}_{2}\right),
\end{eqnarray}
satisfying the commutation relation $[S_i,S_j]=i \epsilon_{ijk} S_k$ with $\epsilon_{ijk}$ being the Levi-Civita symbol. $\hat{H}^R_{\rm int}(t)$  read then: 

\be
\hat{H}^R_{\rm int}(t)=\dfrac{g_{11}}{2} \int_{-\infty}^{\infty}\,d\mathbf{r}\,|\phi_1(\mathbf{r},t)|^4\hat{a}^\dagger_{1}\hat{a}^\dagger_{1}\hat{a}_{1}\hat{a}_{1}+\dfrac{g_{22}}{2} \int_{-\infty}^{\infty}\,d\mathbf{r}\,|\phi_2(\mathbf{r},t)|^4\hat{a}^\dagger_{2}\hat{a}^\dagger_{2}\hat{a}_{2}\hat{a}_{2}+g_{12} \int_{-\infty}^{\infty}\,d\mathbf{r}\,|\phi_1(\mathbf{r},t)|^2.|\phi_2(\mathbf{r},t)|^2\hat{a}^\dagger_{1}\hat{a}^\dagger_{2}\hat{a}_{1}\hat{a}_{2}.
\ee
Using the two relations,

\begin{eqnarray}
\label{eqA_Sz2_N2}
    \hat{N}^2 &=& \hat{a}^\dagger_{1}\hat{a}^\dagger_{1}\hat{a}_{1}\hat{a}_1+\hat{a}^\dagger_{2}\hat{a}^\dagger_{2}\hat{a}_2\hat{a}_2+2\hat{a}^\dagger_{1}\hat{a}_{1}\hat{a}^\dagger_{2}\hat{a}_{2}+\hat{N}\, \nonumber \\
    (2\hat{S}_z)^2 &=& \hat{a}^\dagger_{1}\hat{a}^\dagger_{1}\hat{a}_{1}\hat{a}_1+\hat{a}^\dagger_{2}\hat{a}^\dagger_{2}\hat{a}_2\hat{a}_2-2\hat{a}^\dagger_{1}\hat{a}_{1}\hat{a}^\dagger_{2}\hat{a}_{2}+\hat{N},
\end{eqnarray}

we find 
\begin{eqnarray}
\label{eqA_Sz2_N2_Simplified}
    \hat{a}^\dagger_{1}\hat{a}^\dagger_{1}\hat{a}_{1}\hat{a}_1+\hat{a}^\dagger_{2}\hat{a}^\dagger_{2}\hat{a}_2\hat{a}_2 &=& \dfrac{\hat{N}^2+(2\hat{S}_z)^2}{2}-\hat{N} \equiv 2 (\hat{S}_z)^2, \nonumber \\
    \hat{a}^\dagger_{1}\hat{a}_{1}\hat{a}^\dagger_{2}\hat{a}_{2} &=& \dfrac{\hat{N}^2-(2\hat{S}_z)^2-2\hat{N}}{4} \equiv - (\hat{S}_z)^2.
\end{eqnarray}

Typically one find, 

\be
\hat{H}^R_{\rm int}(t)=\hbar\left[\chi_{11}(t)+\chi_{22}(t)-2\chi_{12}(t)\right](\hat{S}_z)^2,
\ee

where $\chi_{ij}(t)=g_{ij}\int d\mathbf{r} |\Phi_{i}(\mathbf{r},t)|^2|\Phi_{j}(\mathbf{r},t)|^2/2\hbar$. The pre-factor $g_{ij}/2\hbar=\pi\,\hbar\, a_{ij}/m_{ij}$ is associated to the s-wave scattering length $a_{ij}$ with $m_{ij}$ being the reduced mass $m_{ij}=m_i m_j/(m_i+m_j)$. In the case where $g_{11}\approx g_{22} \approx g_{12} \equiv g$ and
$|\phi_1(\mathbf{r},t)|^2=|\phi_2(\mathbf{r},t)|^2$,
$\hat{H}^R_{\rm int}(t)$ simplified to 

\be
\hat{H}^R_{\rm int}(t)=\hbar\left[\chi_S(t)-\chi_C(t)\right]\hat{S}_z^2,
\ee

where $\chi_{S}(t)=g \int_{-\infty}^{\infty}\,d\mathbf{r}\,\left|\Phi(\mathbf{r},t)\right|^4/\hbar$ denotes the self-phase modulation non-linear coefficient ($\left|\Phi_1(\mathbf{r},t)\right|^2=\left|\Phi_2(\mathbf{r},t)\right|^2\equiv \left|\Phi(\mathbf{r},t)\right|^2$) and 
$\chi_{C}(t)=g\int_{-\infty}^{\infty}\,d\mathbf{r}\,\left|\Phi_{1}(\mathbf{r},t)\right|^2 \left|\Phi_{2}(\mathbf{r},t)\right|^2/\hbar$ denotes the cross-phase modulation non-linear coefficient.

\subsection{One-component two external modes Hamiltonian (Bragg)}

Here we consider the one-component non-linear Hamiltonian,

\be
\hat{H}^B_{\rm int}(t)=\dfrac{g}{2} \int_{-\infty}^{\infty}\,d\mathbf{r}\,\hat{\Psi}^\dagger\hat{\Psi}^\dagger\hat{\Psi}\hat{\Psi},
\ee

where the field operators, $\hat{\Psi}=\phi_1(\mathbf{r},t)\, \hat{a}_1+\phi_2(\mathbf{r},t)\,e^{2ikz}\, \hat{a}_2$. As before $\phi_i(\mathbf{r},t)$ denotes the spatial shape of component $\hat{\Psi}_i$ and $ e^{2ikz}$ is the phase imprinted by the two-photon transition. The relations given in Eqs.(\ref{eqA_Sx_Sy_Sz_def}), (\ref{eqA_Sz2_N2}) and (\ref{eqA_Sz2_N2_Simplified}) still hold. In this case the term, $\hat{\Psi}^\dagger\hat{\Psi}^\dagger\hat{\Psi}\hat{\Psi}$, read as:

\begin{eqnarray}
\hat{\Psi}^\dagger\hat{\Psi}^\dagger\hat{\Psi}\hat{\Psi} &=& |\phi_1(\mathbf{r},t)|^4\hat{a}^\dagger_{1}\hat{a}^\dagger_{1}\hat{a}_{1}\hat{a}_{1} + |\phi_2(\mathbf{r},t)|^4\hat{a}^\dagger_{2}\hat{a}^\dagger_{2}\hat{a}_{2}\hat{a}_{2} + 4\,|\phi_1(\mathbf{r},t)|^2.|\phi_2(\mathbf{r},t)|^2\hat{a}^\dagger_{1}\hat{a}^\dagger_{2}\hat{a}_{1}\hat{a}_{2} \nonumber \\
&&+\phi_1(\mathbf{r},t)^2.\phi_2(\mathbf{r},t)^2 e^{4ikz}\hat{a}^\dagger_{1}\hat{a}^\dagger_{1}\hat{a}_{2}\hat{a}_{2}+2|\phi_1(\mathbf{r},t)|^2.\phi_1(\mathbf{r},t).\phi_2(\mathbf{r},t) e^{2ikz}\hat{a}^\dagger_{1}\hat{a}^\dagger_{1}\hat{a}_{1}\hat{a}_{2} \nonumber \\
&&+ \phi_1(\mathbf{r},t)^2.\phi_2(\mathbf{r},t)^2 e^{-4ikz}\hat{a}^\dagger_{2}\hat{a}^\dagger_{2}\hat{a}_{1}\hat{a}_{1}+2.\phi_1(\mathbf{r},t).|\phi_2(\mathbf{r},t)|^2.\phi_2(\mathbf{r},t) e^{-2ikz}\hat{a}^\dagger_{2}\hat{a}^\dagger_{2}\hat{a}_{1}\hat{a}_{2} \nonumber \\
&&+2\phi_1(\mathbf{r},t).|\phi_1(\mathbf{r},t)|^2.\phi_2(\mathbf{r},t)e^{-2ikz}\hat{a}^\dagger_{1}\hat{a}^\dagger_{2}\hat{a}_{1}\hat{a}_{1}+2 \phi_2(\mathbf{r},t)e^{2ikz}.|\phi_2(\mathbf{r},t)|^2.\phi_1(\mathbf{r},t)\hat{a}^\dagger_{1}\hat{a}^\dagger_{2}\hat{a}_{2}\hat{a}_{2}, 
\end{eqnarray}

and after simplification, where we neglect the terms  $\int_{-\infty}^{\infty}\,d\mathbf{r}\, \phi_i(\mathbf{r},t).\phi_j(\mathbf{r},t).\phi_k(\mathbf{r},t).\phi_l(\mathbf{r},t)e^{inkz}\approx 0$, we find:

\be
\hat{H}^B_{\rm int}(t)=\dfrac{g}{2}\int_{-\infty}^{\infty}\,d\mathbf{r}\,\left( |\phi_1(\mathbf{r},t)|^4\hat{a}^\dagger_{1}\hat{a}^\dagger_{1}\hat{a}_{1}\hat{a}_{1}+ |\phi_2(\mathbf{r},t)|^4\hat{a}^\dagger_{2}\hat{a}^\dagger_{2}\hat{a}_{2}\hat{a}_{2}+4|\phi_1(\mathbf{r},t)|^2.|\phi_2(\mathbf{r},t)|^2\hat{a}^\dagger_{1}\hat{a}^\dagger_{2}\hat{a}_{1}\hat{a}_{2}\right).
\ee

In the case where 
$|\phi_1(\mathbf{r},t)|^2=|\phi_2(\mathbf{r},t)|^2$,
$\hat{H}^B_{\rm int}(t)$ simplified to

\be
\hat{H}^B_{\rm int}(t)=\hbar\left[\chi_S(t)-2\,\chi_C(t)\right]\hat{S}_z^2.
\ee

\section{Non-linear readout}
\subsection{Sensitivity Gain}
We consider the initial coherent state
\begin{equation}
\label{eq_CSS_input}
    |\psi_0\rangle = 2^{-S}\sum \limits_{n=0}^{2S}\binom{2S}{n}^{1/2}|S,S-n\rangle,
\end{equation}
where $S=N/2$.

The sensitivity of the interferometer, $\Delta \theta$, is evaluated via the error propagation formula,
\be
\label{eq_App_ErrorPropagator}
(\Delta \theta)^2 = \dfrac{(\Delta S_y)^2_{\rm out}}{( d\langle \hat{S}_y \rangle_{\rm out} / d \theta )^2}\bigg|_{(\theta=0)},
\ee
where the state at the end of the interferometer sequence is given by:

\be
|\Psi_{\rm out}\rangle=\hat{U}^\dagger e^{i(\theta\hat{S}_y+\eta \hat{S}_y^2)}\hat{U}|\Psi_0\rangle,
\ee

where $\eta=\tau_{AI}$.

We define the sensitivity gain in the case of a non-linear readout by $\mathcal{Q}$ such as

\be
\mathcal{Q}^2=\dfrac{1}{N (\Delta \theta)^2}
\ee

\subsection{Case where the phase and non-linear term during the interferometer are considered small}
In the situation where $\theta$ and $\eta$ are small enough to approximate  $e^{i(\theta\hat{S}_y+\eta \hat{S}_y^2)}\approx1+i(\theta\hat{S}_y+\eta \hat{S}_y^2)$ we have,

\begin{align}
\langle\Psi_{\rm out}|\hat{S}_y|\Psi_{\rm out}\rangle&= \langle\Psi_0|\hat{S}_y|\Psi_0\rangle+i\theta\langle\Psi_0|[\hat{S}_y,\hat{U}^\dagger \hat{S}_y\hat{U}]|\Psi_0\rangle+i\eta\langle\Psi_0|[\hat{S}_y,\hat{U}^\dagger \hat{S}_y^2\hat{U}]|\Psi_0\rangle, \\
\dfrac{d\langle \Psi_{\rm out}|\hat{S}_y|\Psi_{\rm out}\rangle}{d\theta}\bigg|_{(\theta=0)}&=i\langle\Psi_0|[\hat{S}_y,\hat{U}^\dagger \hat{S}_y\hat{U}]|\Psi_0\rangle+\eta\langle\Psi_0|\left(\hat{U}^\dagger \hat{S}_y\hat{U}\hat{S}_y\hat{U}^\dagger\hat{S}_y^2\hat{U}+\hat{U}^\dagger\hat{S}_y^2\hat{U}\hat{S}_y\hat{U}^\dagger \hat{S}_y\hat{U}\right) |\Psi_0\rangle \\
\langle\Psi_{\rm out}|\hat{S}_y^2|\Psi_{\rm out}\rangle &= \langle\Psi_0|\hat{S}_y^2|\Psi_0\rangle+i\theta\langle\Psi_0|[\hat{S}_y^2,\hat{U}^\dagger \hat{S}_y\hat{U}]|\Psi_0\rangle+i\eta\langle\Psi_0|[\hat{S}_y^2,\hat{U}^\dagger \hat{S}_y^2\hat{U}]|\Psi_0\rangle.
\end{align}

To perform the calculation of the different terms of the previous equations we use the following relations:

\begin{align*}
\hat{S}_y&=(S_+-S_-)/2i \nonumber\\
\hat{U}^\dagger\hat{S}_+\hat{U}&=\hat{S}_+\hat{W}\nonumber\\
\hat{W}&=\exp[i\tau(2\hat{S}_z+1)]\nonumber\\
\hat{W}\hat{S}_+&=\hat{S}_+\hat{W}\exp(2i\tau)\nonumber\\
\hat{W}^\dagger\hat{S}_+&=\hat{S}_+\hat{W}^\dagger\exp(-2i\tau)\nonumber\\
\hat{V}&=\exp[4i\tau(\hat{S}_z+1)]=\hat{W}^2\exp(2i\tau)\nonumber\\
\hat{V}\hat{S}_+&=\hat{S}_+\hat{V}\exp(4i\tau)\\
\hat{V}^\dagger\hat{S}_+&=\hat{S}_+\hat{V}^\dagger\exp(-4i\tau)\nonumber\\
\hat{S}_y^2&=-\dfrac{1}{4}\left(\hat{S}_+^2+\hat{S}_-^2+2\hat{S}_z-2\hat{S}_+\hat{S}_-\right)\nonumber\\
\hat{U}^\dagger\hat{S}_y\hat{U}&=\dfrac{1}{2i}\left(\hat{S}_+\hat{W}-\hat{W}^\dagger\hat{S}_-\right)\nonumber\\
\hat{U}^\dagger\hat{S}_y^2\hat{U}&=-\dfrac{1}{4}\left(\hat{S}_+^2\hat{V}+\hat{V}^\dagger\hat{S}_-^2+2\hat{S}_z-2\hat{S}_+\hat{S}_-\right).\nonumber
\end{align*}

After simplification the different terms read

\begin{align*}
[\hat{S}_y,\hat{U}^\dagger\hat{S}_y\hat{U}]=&-\dfrac{1}{4}\left(\hat{S}_+^2\hat{W}[1-\exp(2i\tau)]+\hat{S}_+\hat{W}\hat{S}_-[1-\exp(2i\tau)]+2\hat{S}_z\hat{W}-H.C.\right)
\end{align*}

\begin{align*}
[\hat{S}_y^2,\hat{U}^\dagger\hat{S}_y\hat{U}]=&-\dfrac{1}{8i}\left(\hat{S}_+^3\hat{W}[1-\exp(4i\tau)]+\hat{S}_+^2\hat{W}^\dagger\hat{S}_-[\exp(-4i\tau)-1]-2\hat{S}_+(2\hat{S}_z+1)[\hat{W}-\hat{W}^\dagger\exp(-2i\tau)]+H.C\right)
\end{align*}

\begin{align*}
[\hat{S}_y,\hat{U}^\dagger\hat{S}_y^2\hat{U}]=&-\dfrac{1}{8i}\left(\hat{S}_+^3\hat{V}[1-\exp(4i\tau)]+\hat{S}_+^2\hat{V}\hat{S}_-[1-\exp(4i\tau)]+2\hat{S}_+(2\hat{S}_z+1)[V-1]+H.C.\right)
\end{align*}

\begin{align*}
[\hat{S}_y^2,\hat{U}^\dagger\hat{S}_y^2\hat{U}]=&\dfrac{1}{16}\left(\hat{S}_+^4\hat{V}[1-\exp(8i\tau)]-\hat{S}_+^2 \hat{V}\hat{S}_-^2[1-\exp(8i\tau)]+4\hat{S}_z(2\hat{S}_z+1)\hat{V}-H.C.\right) \\
&+\dfrac{1}{2}\left(\hat{S}_+^2\hat{S}_z\hat{V}+\hat{S}_+^2\hat{V}-\hat{S}_+^2\hat{S}_z-\hat{S}_+^2-\hat{S}_+(\hat{S}_z+1)\hat{V}\hat{S}_-\exp(4 i \tau)-H.C.\right)
\end{align*}
\begin{align*}
\hat{U}^\dagger \hat{S}_y\hat{U}\hat{S}_y\hat{U}^\dagger\hat{S}_y^2\hat{U}&+\hat{U}^\dagger\hat{S}_y^2\hat{U}\hat{S}_y\hat{U}^\dagger \hat{S}_y\hat{U}=\dfrac{1}{16}\Big\{\nonumber \\
&+4 \hat{S}_z^2\hat{W}+4\hat{S}_z(2\hat{S}_z+1)\hat{V}\hat{W}^\dagger+\hat{S}_+^4\hat{V}\hat{W}[\exp(8 i \tau)+\exp(6i\tau)]\nonumber \\ 
&+\hat{S}_+^2\hat{S}_z\left(6\hat{V}\hat{W}^\dagger \exp(-4i\tau)+10\hat{W}+2\hat{W}\exp(2i\tau)+2\hat{V}\hat{W}+4\hat{V}\hat{W}\exp(2i\tau)\right)\nonumber \\
&+\hat{S}_+^2\left(8\hat{W}+6\hat{V}\hat{W}^\dagger\exp(-4i\tau)+2\hat{V}\hat{W}\exp(2i\tau)\right) \nonumber \\
&+\hat{S}_+^2\left(\hat{V}^\dagger\hat{W}\left[\exp(2 i\tau)+\exp(-4i\tau)\right]+2\hat{W}\left[\exp(2 i\tau)+\exp(4i\tau)\right]\right)\hat{S}_-^2 \\
&-\hat{S}_+\hat{S}_z\left(6\hat{W}+10\hat{W}\exp(2i\tau)+8\hat{V}\hat{W}^\dagger\exp(2i\tau)\right)\hat{S}_-\nonumber \\
&-\hat{S}_+^3\left(2\hat{W}\left[\exp(4i\tau)+\exp(2i\tau)\right]+\hat{V}\hat{W}\left[\exp(8i\tau)+\exp(6i\tau)\right]+\hat{V}\hat{W}^\dagger\left[\exp(-2i\tau)+\exp(4i\tau)\right]\right)\hat{S}_-\nonumber\\
&-\hat{S}_+\left(2\hat{W}+6\hat{W}\exp(2i\tau)+8\hat{V}\hat{W}^\dagger \exp(2i\tau)\right)\hat{S}_-+H.C.\Big\}\nonumber
\end{align*}

Using the results derived in Appendix\,\ref{help} we find,

\begin{subequations}
\begin{align}
\langle\Psi_{\rm out}|[\hat{S}_y,\hat{U}^\dagger\hat{S}_y\hat{U}]|\Psi_{\rm out}\rangle&=i S(2S-1)\sin(\tau)\cos(\tau)^{2S-2} \\
\langle\Psi_{\rm out}|[\hat{S}_y^2,\hat{U}^\dagger\hat{S}_y\hat{U}]|\Psi_{\rm out}\rangle&=-2iS(2S-1)\sin(\tau)^2\cos(\tau)^{2S-1}\\ 
\langle\Psi_{\rm out}|[\hat{S}_y,\hat{U}^\dagger\hat{S}_y^2\hat{U}]|\Psi_{\rm out}\rangle&=0\\
\langle\Psi_{\rm out}|[\hat{S}_y^2,\hat{U}^\dagger\hat{S}_y^2\hat{U}]|\Psi_{\rm out}\rangle&=i\dfrac{S}{2}\Big[(3+4S(S-1)+2(S+1)\cos(4\tau)+\cos(8\tau))\sin(2\tau)\cos(2\tau)^{2S-3}\\
&-(2S+1)\sin(4\tau)\Big]\nonumber\\
\langle\Psi_{\rm out}|\hat{U}^\dagger \hat{S}_y\hat{U}\hat{S}_y\hat{U}^\dagger\hat{S}_y^2\hat{U}&+\hat{U}^\dagger\hat{S}_y^2\hat{U}\hat{S}_y\hat{U}^\dagger \hat{S}_y\hat{U}|\Psi_{\rm out}\rangle=\nonumber\\
&-\dfrac{S}{8}(2S-1)\cos(3\tau)^{2S-3}\left[(2S-1)\cos(2\tau)+S\cos(4\tau)\right] \\
&+\dfrac{S}{8}\cos(\tau)^{2S-3}\left[1-4S+8S^2+(-1+2S+4S^2)\cos(2\tau)-3(S-1)(2S-1)\cos(4\tau)\right] \nonumber
\end{align}
\end{subequations}

\subsection{Sensitivity Gain in the perturbative regime}

The sensitivity gain read,

\be
\mathcal{Q}^2=(2S-1)^2\sin(\tau)^2\cos(\tau)^{4S-4} +\mathcal{A} \,\eta +\mathcal{O}(\eta^2)
\ee
with

\begin{subequations}
\begin{align}
\mathcal{A}&=\dfrac{1}{4}(2S-1)^2\sin(\tau)\cos(\tau)^{2S-2}\cos(3\tau)^{2S-3}\left[(2S-1)\cos(2\tau)+S\cos(4\tau)\right] \\
&+2(2S-1)^2\sin(\tau)^3\cos(\tau)^{4S-3}\left[-2(2S+1)\cos(2\tau)+\cos(2\tau)^{2S-3}\left(3+4S(S-1)+2(S+1)\cos(4\tau)+\cos(8\tau)\right)\right] \nonumber \\
&+\dfrac{1}{4}(2S-1)\sin(\tau)\cos(\tau)^{4S-5}\left[-1+4S-8S^2+(1-2S(2S+1))\cos(2\tau)+(3-9S+6S^2)\cos(4\tau)\right]
\end{align}
\end{subequations}

To first order in $\tau$ and $\eta$ one finds,

\be
\mathcal{Q}^2(\tau,\eta)=\frac{1}{4}\left[(4\tau-3\eta)S-(2\tau-\eta)\right]^2+\mathcal{O}(\tau^3)+\mathcal{O}(\eta^3).
\ee

\subsection{Help for the calculation}
\label{help}

\begin{align*} 
&\langle\Psi_0|\hat{S}_+\hat{W}|\Psi_0\rangle=S\cos(\tau)^{2S-1} 
\\&\langle\Psi_0|\hat{S}_+^2\hat{W}|\Psi_0\rangle=\dfrac{S}{2}(2S-1)\cos(\tau)^{2S-2}\exp(-i\tau)
\\&\langle\Psi_0|\hat{S}_+^3\hat{W}|\Psi_0\rangle=\dfrac{S}{2}(2S-1)(S-1)\cos(\tau)^{2S-3}\exp(-2i\tau)
\end{align*}
\begin{align*} 
&\langle\Psi_0|\hat{S}_+\hat{W}^\dagger|\Psi_0\rangle=\langle\Psi_0|\hat{S}_+\hat{W}|\Psi_0\rangle
\\&\langle\Psi_0|\hat{S}_+\hat{W}\hat{S}_z|\Psi_0\rangle=-\dfrac{S}{2}\cos(\tau)^{2S-1}+i\dfrac{S}{2}(2S-1)\sin(\tau)\cos(\tau)^{2S-2}
\\&\langle\Psi_0|\hat{S}_+\hat{W}^\dagger\hat{S}_z|\Psi_0\rangle=-\dfrac{S}{2}\cos(\tau)^{2S-1}-i\dfrac{S}{2}(2S-1)\sin(\tau)\cos(\tau)^{2S-2}
\end{align*}
\begin{align*} 
&\langle\Psi_0|\hat{S}_z\hat{W}|\Psi_0\rangle=iS\sin(\tau)\cos(\tau)^{2S-1}\exp(i\tau)
\\&\langle\Psi_0|\hat{S}_+\hat{W}\hat{S}_-|\Psi_0\rangle=S\cos(\tau)^{2S-1}+\dfrac{S}{2}(2S-1)\cos(\tau)^{2S-2}\exp(-i\tau)
\\&\langle\Psi_0|\hat{S}_+^2\hat{W}^\dagger\hat{S}_-|\Psi_0\rangle=S(2S-1)\cos(\tau)^{2S-2}\exp(i\tau)+\dfrac{S}{2}(2S-1)(S-1)\cos(\tau)^{2S-3}\exp(2i\tau)
\end{align*}
\begin{align*} 
&\langle\Psi_0|\hat{S}_z^2\hat{W}|\Psi_0\rangle=\dfrac{S}{2}\cos(\tau)^{2S-2}(1-S+S\cos(2\tau))\exp(i\tau)
\\&\langle\Psi_0|\hat{S}_+^2\hat{S}_z\hat{W}|\Psi_0\rangle=\dfrac{S}{4}(2S-1)\cos(\tau)^{2S-3}\left[(S-2)-S\exp(-2i\tau)\right]
\\&\langle\Psi_0|\hat{S}_+\hat{S}_z\hat{W}\hat{S}_-|\Psi_0\rangle=\dfrac{S}{4}\left[1+(S-1)\exp(2i\tau)-2S^2\exp(-2i\tau)+S(2S-5)\right]\cos(\tau)^{2S-3}
\end{align*}
\begin{align*} 
&\langle\Psi_0|\hat{S}_+^2\hat{S}_z\hat{V}\hat{W}|\Psi_0\rangle=\dfrac{S}{4}(2S-1)\left[(S-2)\exp(2i\tau)-S\exp(-4i\tau)\right]\cos(3\tau)^{2S-3}
\\&\langle\Psi_0|\hat{S}_+^4\hat{V}\hat{W}|\Psi_0\rangle=\dfrac{S}{4}(2S-1)(S-1)(2S-3)\cos(3\tau)^{2S-4}\exp(-7i\tau)
\\&\langle\Psi_0|\hat{S}_+^2\hat{V}\hat{W}^\dagger\hat{S}_-^2|\Psi_0\rangle=\dfrac{S}{4}(2S-1)\left[\exp(3i\tau)+S(2S-1)\exp(-i\tau)+(4S-2)\exp(i\tau)\right]\cos(\tau)^{2S-4}
\end{align*}
\begin{align*} 
&\langle\Psi_0|\hat{S}_+^3\hat{V}\hat{W}\hat{S}_-|\Psi_0\rangle=\dfrac{S}{4}(2S-1)(S-1)\left[3\exp(i\tau)+2S\exp(-7i\tau)\right]\cos(3\tau)^{2S-4}
\\&\langle\Psi_0|\hat{S}_+^3\hat{V}\hat{W}^\dagger\hat{S}_-|\Psi_0\rangle=\dfrac{S}{4}(2S-1)(S-1)\left[3\exp(i\tau)+2S\exp(-i\tau)\right]\cos(\tau)^{2S-4}
\\&\langle\Psi_0|\hat{S}_+^2\hat{S}_z\hat{V}\hat{W}^\dagger|\Psi_0\rangle=\dfrac{S}{4}(2S-1)[(S-2)\exp(2i\tau)-S]\cos(\tau)^{2S-3}
\\&\langle\Psi_0|\hat{S}_+^2\hat{V}\hat{W}^\dagger|\Psi_0\rangle=\dfrac{S}{2}(2S-1)\cos(\tau)^{2S-2}\exp(i\tau)
\end{align*}
\begin{align*} 
&\langle\Psi_0|\hat{S}_+\hat{S}_z\hat{V}^\dagger\hat{W}\hat{S}_-|\Psi_0\rangle=\dfrac{S}{4}\left[-1+S+\exp(2i\tau)\left(1-S\{5+2S(-1+\exp(2i\tau)))\}\right)\right]\cos(\tau)^{2S-3}\exp(-4i\tau)
\\&\langle\Psi_0|\hat{S}_+\hat{V}^\dagger\hat{W}\hat{S}_-|\Psi_0\rangle=\dfrac{S}{2}[2S\exp(-i\tau)+\exp(-3i\tau)]\cos(\tau)^{2S-2}
\\&\langle\Psi_0|\hat{S}_z^2\hat{V}^\dagger\hat{W}|\Psi_0\rangle=\dfrac{S}{2}[1-S+S\cos(2\tau)]\cos(\tau)^{2S-2}\exp(-3i\tau)
\end{align*}
\begin{align*} 
&\langle\Psi_0|\hat{S}_z\hat{V}^\dagger\hat{W}|\Psi_0\rangle=-i S \sin(\tau) \cos(\tau)^{2S-1}\exp(-3i\tau)
\\&\langle\Psi_0|\hat{S}_+^2\hat{V}\hat{W}|\Psi_0\rangle=\dfrac{S}{2}(2S-1)\cos(3\tau)^{2S-2}\exp(-i\tau)
\\&\langle\Psi_0|\hat{S}_+^3\hat{W}\hat{S}_-|\Psi_0\rangle=\dfrac{S}{4}(2S-1)(S-1)[2S\exp(-3i\tau)+3\exp(-i\tau)]\cos(\tau)^{2S-4}
\\&\langle\Psi_0|\hat{S}_+^2\hat{W}\hat{S}_-^2|\Psi_0\rangle=\dfrac{S}{4}(2S-1)\left[\exp(i\tau)+S(2S-1)\exp(-3i\tau)+(4S-2)\exp(-i\tau)\right]\cos(\tau)^{2S-4}
\end{align*}
\begin{align*} 
&\langle\Psi_0|\hat{S}_+|\Psi_0\rangle=S
\\&\langle\Psi_0|\hat{S}_+^2|\Psi_0\rangle=\dfrac{S}{2}(2S-1)
\\&\langle\Psi_0|\hat{S}_+\hat{V}|\Psi_0\rangle=S\cos(2\tau)^{2S-1}\exp(2i\tau)
\\&\langle\Psi_0|\hat{S}_+^2\hat{V}|\Psi_0\rangle=\dfrac{S}{2}(2S-1)\cos(2\tau)^{2S-2}
\\&\langle\Psi_0|\hat{S}_+^3\hat{V}|\Psi_0\rangle=\dfrac{S}{2}(2S-1)(S-1)\cos(2\tau)^{2S-3}\exp(-2i\tau)
\end{align*}
\begin{align*} 
&\langle\Psi_0|\hat{S}_+^4\hat{V}|\Psi_0\rangle=\dfrac{S}{4}(2S-1)(S-1)(2S-3)\cos(2\tau)^{2S-4}\exp(-4i\tau)
\\&\langle\Psi_0|\hat{S}_+\hat{S}_z|\Psi_0\rangle=-\dfrac{S}{2}
\\&\langle\Psi_0|\hat{S}_+^2\hat{S}_z|\Psi_0\rangle=-\dfrac{S}{2}(2S-1)
\\&\langle\Psi_0|\hat{S}_z\hat{V}|\Psi_0\rangle=iS\sin(2\tau)\cos(2\tau)^{2S-1}\exp(4i\tau)
\end{align*}
\begin{align*} 
&\langle\Psi_0|\hat{S}_z^2\hat{V}|\Psi_0\rangle=\dfrac{S}{2}[S\cos(4\tau)-(S-1)]\exp(4i\tau)\cos(2\tau)^{2S-2}
\\&\langle\Psi_0|\hat{S}_+\hat{S}_z\hat{V}|\Psi_0\rangle=-\dfrac{S}{2}\cos(2\tau)^{2S-2}\exp(4i\tau)+iS^2\sin(2\tau)\cos(2\tau)^{2S-2}\exp(2i\tau)
\\&\langle\Psi_0|\hat{S}_+^2\hat{S}_z\hat{V}|\Psi_0\rangle=-\dfrac{S}{2}(2S-1)\cos(2\tau)^{2S-2}+i\dfrac{S}{2}(2S-1)(S-1)\sin(2\tau)\cos(2\tau)^{2S-3}
\end{align*}
\begin{align*} 
&\langle\Psi_0|\hat{S}_+^2\hat{V}\hat{S}_-|\Psi_0\rangle=\dfrac{S}{2}(S+\exp(4i\tau))(2S-1)\cos(2\tau)^{2S-3}\exp(-2i\tau)
\\&\langle\Psi_0|\hat{S}_+^2\hat{V}\hat{S}_-^2|\Psi_0\rangle=\dfrac{S}{4}(2S-1)[\exp(4i\tau)+2(2S-1)+S(2S-1)\exp(-4i\tau)]\cos(2\tau)^{2S-4}
\\&\langle\Psi_0|\hat{S}_+\hat{S}_z\hat{S}_-|\Psi_0\rangle=-S^2
\end{align*}
\begin{align*} 
&\langle\Psi_0|\hat{S}_+\hat{S}_z\hat{V}\hat{S}_-|\Psi_0\rangle=\dfrac{S}{4}[(S-1)\exp(6i\tau)+(1-5S)\exp(2i\tau)+4iS^2\sin(2\tau)]\cos(2\tau)^{2S-3}
\\&\langle\Psi_0|\hat{S}_+\hat{V}\hat{S}_-|\Psi_0\rangle=\dfrac{S}{2}(2S-1)\cos(2\tau)^{2S-2}+S\cos(2\tau)^{2S-1}\exp(2i\tau)
\end{align*}


\begin{thebibliography}{10}
\makeatletter
\providecommand \@ifxundefined [1]{%
 \@ifx{#1\undefined}
}%
\providecommand \@ifnum [1]{%
 \ifnum #1\expandafter \@firstoftwo
 \else \expandafter \@secondoftwo
 \fi
}%
\providecommand \@ifx [1]{%
 \ifx #1\expandafter \@firstoftwo
 \else \expandafter \@secondoftwo
 \fi
}%
\providecommand \natexlab [1]{#1}%
\providecommand \enquote  [1]{``#1''}%
\providecommand \bibnamefont  [1]{#1}%
\providecommand \bibfnamefont [1]{#1}%
\providecommand \citenamefont [1]{#1}%
\providecommand \href@noop [0]{\@secondoftwo}%
\providecommand \href [0]{\begingroup \@sanitize@url \@href}%
\providecommand \@href[1]{\@@startlink{#1}\@@href}%
\providecommand \@@href[1]{\endgroup#1\@@endlink}%
\providecommand \@sanitize@url [0]{\catcode `\\12\catcode `\$12\catcode
  `\&12\catcode `\#12\catcode `\^12\catcode `\_12\catcode `\%12\relax}%
\providecommand \@@startlink[1]{}%
\providecommand \@@endlink[0]{}%
\providecommand \url  [0]{\begingroup\@sanitize@url \@url }%
\providecommand \@url [1]{\endgroup\@href {#1}{\urlprefix }}%
\providecommand \urlprefix  [0]{URL }%
\providecommand \Eprint [0]{\href }%
\providecommand \doibase [0]{http://dx.doi.org/}%
\providecommand \selectlanguage [0]{\@gobble}%
\providecommand \bibinfo  [0]{\@secondoftwo}%
\providecommand \bibfield  [0]{\@secondoftwo}%
\providecommand \translation [1]{[#1]}%
\providecommand \BibitemOpen [0]{}%
\providecommand \bibitemStop [0]{}%
\providecommand \bibitemNoStop [0]{.\EOS\space}%
\providecommand \EOS [0]{\spacefactor3000\relax}%
\providecommand \BibitemShut  [1]{\csname bibitem#1\endcsname}%
\let\auto@bib@innerbib\@empty

\bibitem{Berman1997}
P.~R. Berman.
{\it Atom Interferometry}.
(Academic Press, San Diego, 1997).

\bibitem{Cronin2009}
A.~D. Cronin, J. Schmiedmayer, and D.~E. Pritchard,
\newblock Optics and interferometry with atoms and molecules,
\newblock {\it Reviews of Modern Physics}, {\bf 81}, 1051 (2009).

\bibitem{Tino2014}
G.~M. Tino and M. A. Kasevich,
{\it Atom Interferometry: Proceedings of the {{International School}}
  of {{Physics}} "{{Enrico Fermi}}", Course 188}
(Societ\'a Italiana di Fisica, Bologna, 2014).

\bibitem{Safronova2018}
M.~S. Safronova, D.~Budker, D.~DeMille, D. F.~J. Kimball,
  A.~Derevianko, and C.~W. Clark,
\newblock Search for new physics with atoms and molecules,
{\it Rev. Mod. Phys.} {\bf 90}, 025008 (2018).

\bibitem{Bongs2019}
K. Bongs, M. Holynski, J. Vovrosh, P. Bouyer, G. Condon, E. Rasel, C. Schubert, W.~P. Schleich, and A. Roura,
\newblock Taking atom interferometric quantum sensors from the laboratory to
  real-world applications,
\newblock {\it Nature Reviews Physics} {\bf 1}, 731 (2019).

\bibitem{arxiv_Geiger_2020}
R.~Geiger, A.~Landragin, S.~Merlet, and F.~Pereira Dos~Santos,
\newblock High-accuracy inertial measurements with cold-atom sensors, 
{\it AVS Quantum Sci.} {\bf 2}, 024702 (2020).

\bibitem{PRL_Pezze_2009}
L.~Pezz\`e and A.~Smerzi, 
Entanglement, nonlinear dynamics, and the Heisenberg limit,
{\it Phys. Rev. Lett.} {\bf 102}, 100401 (2009).

\bibitem{MRP_Pezze_2018}
L.~Pezz\`e, A.~Smerzi, M.~K. Oberthaler, R.~Schmied, and P.~Treutlein, 
Quantum metrology with nonclassical states of atomic ensembles,
\newblock {\it Rev. Mod. Phys.} {\bf 90}, 035005 (2018).

\bibitem{Nature_Gross_2010}
C.~Gross, T.~Zibold, E.~Nicklas, J.~Estève, and M.~K. Oberthaler,
\newblock Nonlinear atom interferometer surpasses classical precision limit,
{\it Nature} {464}, 1165 (2010).

\bibitem{Nature_Riedel_2010}
M.~F. Riedel, P. B\"ohi, Y. Li, T.~W. H\"ansch, A. Sinatra, and P. Treutlein,
\newblock Atom-chip-based generation of entanglement for quantum metrology.
{\it Nature} {464}, 1170 (2010).

\bibitem{Science_Lucke_2011}
B.~L{\"u}cke, et al.,
\newblock Twin matter waves for interferometry beyond the classical limit,
\newblock {\it Science} {\bf 334}, 773 (2011).

\bibitem{Nature_Hosten_2016}
O.~Hosten, N.~J. Engelsen, R.~Krishnakumar, and M.~A. Kasevich,
\newblock Measurement noise 100 times lower than the quantum-projection limit
  using entangled atoms,
\newblock {\it Nature} {\bf 529}, 505 (2016).

\bibitem{BohnetNATPHOT2014}
J. G. Bohnet et al., 
Reduced spin measurement back-action for a phase sensitivity ten times beyond the standard quantum limit, {\it Nat. Phot.} {\bf 8}, 731 (2014). 

\bibitem{Louchet-ChauvetNJP2010}
A. Louchet-Chauvet, J. Appel, J.J. Renema, D. Oblak, N. Kjaergaard, and E. S. Polzik, Entanglement-assisted atomic clock beyond the projection noise limit, {\it New J. Phys.} {\bf 12}, 065032 (2010). 

\bibitem{LerouxPRL2010}
I. D. Leroux, M. H. Schleier-Smith, and V. Vuletić, Orientation-dependent entanglement lifetime in a squeezed atomic clock, {\it Phys. Rev. Lett.} {\bf 104}, 250801 (2010).

\bibitem{PRL_Kruse_2016}
I.~Kruse, K.~Lange, J.~Peise, B.~L\"ucke, L.~Pezz\`e, J.~Arlt, W.~Ertmer,
  C.~Lisdat, L.~Santos, A.~Smerzi, and C.~Klempt,
\newblock Improvement of an atomic clock using squeezed vacuum,
\newblock {\it Phys. Rev. Lett.} {117}, 143004 (2016).

\bibitem{PedrozoNATURE2020}
E. Pedrozo-Peñafiel, S. Colombo, C. Shu, A. F. Adiyatullin, Z. Li, E. Mendez, B. Braverman, A. Kawasaki, D. Akamatsu, Y. Xiao, V. Vuleti\'c,
Entanglement on an optical atomic-clock transition,
{\it Nature} {\bf 588}, 414 (2020).

\bibitem{AndrePRL2004}
A. André, A. S. S\o rensen, and M. D. Lukin, Stability of atomic clocks based onentangled atoms,
{\it Phys. Rev. Lett.} {\bf 92}, 230801 (2004)

\bibitem{SchulteNATCOMM2020}
M. Schulte, C. Lisdat, P. O. Schmidt, U. Sterr, and K. Hammerer,
Prospects and challenges for squeezing-enhanced optical atomic clocks,
{\it Nat. Comm.} {\bf 11}, 1 (2020).

\bibitem{PRL_Pezze_2020}
L. Pezz\`e and A. Smerzi,
\newblock Heisenberg-limited noisy atomic clock using a hybrid coherent and
  squeezed state protocol,
\newblock {\it Phys. Rev. Lett.} {125}, 210503 (2020).





\bibitem{PRA_Petersen_2005}
Vivi Petersen, Lars~Bojer Madsen, and Klaus M\o{}lmer.
\newblock Magnetometry with entangled atomic samples.
\newblock {\it Phys. Rev. A}, 71:012312, Jan 2005.

\bibitem{PRL_Wasilewski_2010}
W.~Wasilewski, K.~Jensen, H.~Krauter, J.~J. Renema, M.~V. Balabas, and E.~S.
  Polzik.
\newblock Quantum noise limited and entanglement-assisted magnetometry.
\newblock {\it Phys. Rev. Lett.}, 104:133601, Mar 2010.

\bibitem{SewellPRL2012}
R. J. Sewell, M. Koschorreck, M. Napolitano, B. Dubost, N. Behbood, and M. W. Mitchell,
Magnetic Sensitivity Beyond the Projection Noise Limit by Spin Squeezing,
{\it Phys. Rev. Lett.} {\bf 109}, 253605 (2012).

\bibitem{OckeloenPRL2013}
C. F. Ockeloen, R. Schmied, M. F. Riedel, and P. Treutlein, Quantum metrology with a scanning probe atom interferometer, 
{\it Phys. Rev. Lett.} {\bf 111}, 143001 (2014).

\bibitem{MuesselPRA2014}
W. Muessel, H. Strobel, D. Linnemann, D. B. Hume, and M. K. Oberthaler, 
\newblock Scalable spin squeezing for quantum-enhanced magnetometry with Bose-Einstein condensates, 
\newblock {\it Phys. Rev. Lett.} {\bf 113}, 103004 (2014).

\bibitem{PRX_Ruster_2017}
T.~Ruster, H.~Kaufmann, M.~A. Luda, V.~Kaushal, C.~T. Schmiegelow,
  F.~Schmidt-Kaler, and U.~G. Poschinger.
\newblock Entanglement-based dc magnetometry with separated ions.
\newblock {\it Phys. Rev. X}  {\bf 7}, 031050 (2017).

\bibitem{PRL_Haine_2013}
S.~A. Haine,
\newblock Information-recycling beam splitters for quantum enhanced atom
  interferometry,
\newblock {\it Phys. Rev. Lett.} {\bf 110}, 053002 (2013).

\bibitem{PRL_Hamilton_2015}
P. Hamilton, M. Jaffe, J.~M. Brown, L. Maisenbacher, B. Estey,
  and H. M\"uller,
\newblock Atom interferometry in an optical cavity,
\newblock {\it Phys. Rev. Lett.} {\bf 114}, 100405 (2015).

\bibitem{PRL_Salvi_2018}
L. Salvi, N. Poli, V. Vuleti\ifmmode~\acute{c}\else \'{c}\fi{},
  and G.~M. Tino,
\newblock Squeezing on momentum states for atom interferometry,
\newblock {\it Phys. Rev. Lett.} {\bf 120}, 033601 (2018).

\bibitem{QST_shankar_2019}
A. Shankar, L. Salvi, M.~L. Chiofalo, N. Poli, and M.~J. Holland,
\newblock Squeezed state metrology with {Bragg} interferometers operating in a
  cavity,
\newblock {\it Quantum Science and Technology} {\bf 4}, 045010 (2019).

\bibitem{PRL_Szigeti_2020}
S.~S. Szigeti, S.~P. Nolan, J.~D. Close, and S.~A. Haine,
\newblock High-precision quantum-enhanced gravimetry with a Bose-Einstein
  condensate,
\newblock {\it Phys. Rev. Lett.} {\bf 125}, 100402 (2020).

\bibitem{arxiv_Anders_2020}
F.~Anders, et al.,
\newblock Momentum entanglement for atom interferometry
, arXiv:2010.15796.

\bibitem{NJP_Szigeti_2012}
S.~S. Szigeti, J.~D. Debs, J.~J. Hope, N.~P. Robins and J.~D. Close,
\newblock Why momentum width matters for atom interferometry with Bragg pulses,
\newblock {\it New Journal of Physics} {\bf 14}, 023009 (2012).

\bibitem{Nature_Sorensen_2001}
A.~S\o rensen, L.-M. Duan, J.~I. Cirac, and P.~Zoller,
\newblock Many-particle entanglement with bose–einstein condensates,
\newblock {\it Nature} {\bf 409}, 63 (2001).

\bibitem{PRA_Kitagawa_1993}
M. Kitagawa and M. Ueda,
\newblock Squeezed spin states,
\newblock {\it Phys. Rev. A} {\bf 47}, 5138 (1993).

\bibitem{OptLettChu1986}
S.~Chu, J.~E. Bjorkholm, A.~Ashkin, J.~P. Gordon, and L.~W. Hollberg,
Proposal for
  optically cooling atoms to temperatures of the order of 10$^{-6}$ K,
\newblock {\it Opt. Lett.} {\bf 11}, 73 (1986).

\bibitem{PRLAmmann1997}
H.~Ammann and N.~Christensen, 
Delta
  Kick Cooling: A New Method for Cooling Atoms, 
{\it Phys. Rev. Lett.} {\bf 78}, 2088 (1997).

\bibitem{Muntinga2013}
H.~M{\"u}ntinga, et al., 
Interferometry with Bose-Einstein Condensates in Microgravity, 
{\it Physical Review Letters} {\bf 110}, 093602 (2013).

\bibitem{PRLKovachy2015}
T.~Kovachy, J.~M. Hogan, A.~Sugarbaker, S.~M. Dickerson, C.~A. Donnelly,
  C.~Overstreet, and M.~A. Kasevich,
Matter
  Wave Lensing to Picokelvin Temperatures,
{\it Phys. Rev. Lett.} {\bf 114}, 143004 (2015).

\bibitem{NJP_Corgier_2020}
R. Corgier, S. Loriani, H. Ahlers, K. Posso-Trujillo, C.
  Schubert, E.~M. Rasel, E. Charron, and N. Gaaloul,
\newblock Interacting quantum mixtures for precision atom interferometry,
\newblock {\it New Journal of Physics}, {\bf 22}, 123008 (2020).

\bibitem{LeibfriedNATURE2005}
D. Leibfried, et al., 
Creation of a six-atom ‘Schr\"odinger cat’ state,
{\it Nature} {\bf 438}, 639 (2005).

\bibitem{PRL_Davis_2016}
E. Davis, G. Bentsen, and M. Schleier-Smith.
\newblock Approaching the Heisenberg limit without single-particle detection,
{\it Phys. Rev. Lett.} {\bf 116}, 053601 (2016).

\bibitem{PRL_Frowis_2016}
F. Fr\"owis, P. Sekatski, and W. D\"ur.
\newblock Detecting large quantum Fisher information with finite measurement
  precision,
  {\it Phys. Rev. Lett.} {\bf 116}, 090801 (2016).

\bibitem{PRA_Macri_2016}
T. Macr\`{\i}, A. Smerzi, and L. Pezz\`e,
\newblock Loschmidt echo for quantum metrology,
{\it Phys. Rev. A} {\bf 94}, 010102 (2016).

\bibitem{Science_Hosten_2016}
O.~Hosten, R.~Krishnakumar, N.~J. Engelsen, and M.~A. Kasevich,
\newblock Quantum phase magnification,
{\it Science} {\bf 352}, 1552 (2016).

\bibitem{NolanPRL2017}
S. P. Nolan, S. S. Szigeti, and S. A. Haine, Optimal and robust quantum metrology using interaction-based readouts,
{\it Phys. Rev. Lett.} {\bf 119}, 193601 (2017).

\bibitem{AndersPRA2018}
F. Anders, L. Pezz\'e, A. Smerzi, and C. Klempt, Phase magnification by two-axis countertwisting for detection-noise robust interferometry, 
{\it Phys. Rev. A} {\bf 97}, 043813 (2018).

\bibitem{Schulte_2020}
M. Schulte, V.~J. Mart{\'{i}}nez-Lahuerta, M.~S. Scharnagl, and K. Hammerer,
\newblock Ramsey interferometry with generalized one-axis twisting echoes,
\newblock {\it {Quantum}}, 4:268, May 2020.


\bibitem{Raman_vs_Bragg}
Raman pulses create a superposition between two internal states of an atom:
  $|g\rangle\equiv|1,0\rangle$ and $|e\rangle\equiv|2,2\hbar k_L\rangle$, where
  the atom in the internal state 2 acquires a momentum kick $2\hbar k_L$, $k_L$
  being the wave-vector of the laser. Conversely, Bragg pulses create a
  superposition between two momentum states, $|g\rangle\equiv|1,0\rangle$ and
  $|e\rangle\equiv|1,2\hbar k_L\rangle$, where the internal state is unchanged.

\bibitem{Approximation2}
For sake of simplicity we introduce the SU(2) pseudo-spin operators of the
  Lie's algebra, $\hat{S}_x =
  \left(\hat{a}^\dagger_{g}\hat{a}_{e}+\hat{a}_{e}^\dagger\hat{a}_{g}\right)/2$,
  $\hat{S}_y =
  \left(\hat{a}_{g}^\dagger\hat{a}_{e}-\hat{a}_{e}^\dagger\hat{a}_{g}\right)/2i$
  and $\hat{S}_z
  =\left(\hat{a}_{g}^\dagger\hat{a}_{g}-\hat{a}_{e}^\dagger\hat{a}_{e}\right)/2$
  satisfying the commutation relation $[S_i,S_j]=i \epsilon_{ijk} S_k$ with
  $\epsilon_{ijk}$ the Levi-Civita symbol.

\bibitem{supp}
See Supplementary Material.

\bibitem{Approximation1}
This approximation for instance describes the case of Rubidium-87 species in
  the internal states: $|g\rangle\equiv|F=1,m_F=-1\rangle$ and
  $|e\rangle\equiv|F=2,m_F=1\rangle$ where $a_{11} = 100.4\,a_0$ , $a_{12} =
  97.7\,a_0$ and $a_{22} = 95.0\,a_0$ with $a_0$ the Bohr radius.

\bibitem{PRA_Guan_2020}
Q.~Guan, T.~M. Bersano, S.~Mossman, P.~Engels, and D.~Blume,
\newblock Rabi oscillations and Ramsey-type pulses in ultracold bosons: role of
  interactions,
\newblock {\it Phys. Rev. A} {\bf 101}, 063620 (2020).

\bibitem{arxiv_Corgier_2020}
R.~Corgier, L.~Pezz\'e, and A.~Smerzi,
\newblock Non-linear bragg trap interferometer, arXiv:2012.05792. 

\bibitem{Delta_kick_analogy}
This atomic lens is comparable to delta-kick collimation
  realizations~\cite{OptLettChu1986,PRLAmmann1997,Muntinga2013,PRLKovachy2015,NJP_Corgier_2020}
  with the difference that it targets a focusing of the atomic ensemble and not
  a collimation. The experimental implementation of one or the other regime
  follows controlling the duration of the potential flashing.

\bibitem{PRL_Roura_2017}
A. Roura,
\newblock Circumventing heisenberg's uncertainty principle in atom
  interferometry tests of the equivalence principle,
\newblock {\it Phys. Rev. Lett.} {\bf 118}, 160401 (2017).

\bibitem{PRL_DAmico_2017}
G.~D'Amico, G.~Rosi, S.~Zhan, L.~Cacciapuoti, M.~Fattori, and G.~M. Tino,
\newblock Canceling the gravity gradient phase shift in atom interferometry,
\newblock {\it Phys. Rev. Lett.} {\bf 119}, 253201 (2017).

\bibitem{PRL_Overstreet_2018}
C. Overstreet, P. Asenbaum, T. Kovachy, R. Notermans, J.~M. Hogan,
  and M.~A. Kasevich,
\newblock Effective inertial frame in an atom interferometric test of the
  equivalence principle,
\newblock {\it Phys. Rev. Lett.} {\bf 120}, 183604 (2018).

\bibitem{Parameters}
The calculation are shown in the case of a $^{87}Rb$ BEC of $N=10^6$ atoms
  generated in a $2\pi\{150,150,50\}$\,Hz harmonic trap. After creation the BEC
  is released from the trap for $T_0=5$ms to insure low interacting ensemble.

\bibitem{PRA_Wineland_1994}
D.~J. Wineland, J.~J. Bollinger, W.~M. Itano, and D.~J. Heinzen,
\newblock Squeezed atomic states and projection noise in spectroscopy,
\newblock {\it Phys. Rev. A} {\bf 50}, 67 (1994).

\bibitem{Extra_phase}
We neglect here the phase shift that accumulates during state preparation. This
  is justified if the corresponding time is short compared to that of the
  interferometer sequence, $T_{\tau},T_{-\tau}\ll
  T_{\theta}$~\cite{PRL_Szigeti_2020}.

\bibitem{Parameters2}
The calculation have been made for $T_0=5$\,ms, $T_\tau=5$\,ms, $\Delta
  t_1=0.6$\,ms, $T'_\tau=0$, $T_{-\tau}=1$\,ms and $T'_{-\tau}=2$\,ms for
  (c) and $T'_{-\tau}=10$\,ms for (d). In
  this configuration the clouds are diluted at each laser pulse to guarantee
  ideal beam-splitters or mirrors pulse efficiencies.

\bibitem{GarttnerNATPHYS}
M. G\"arttner, J. G. Bohnet, A. Safavi-Naini, M. L. Wall, J. J. Bollinger, and A. M. Rey,
Measuring out-of-time-order correlations and multiple quantum spectra in a trapped-ion quantum magnet,
{\it Nat. Phys.} {\bf 13}, 781 (2017).

\bibitem{PRL_Sinatra_2008}
Y. Li, Y. Castin and A. Sinatra,
\newblock Optimum Spin Squeezing in Bose-Einstein Condensates with Particle Losses,
\newblock {\it Phys. Rev. Lett.}, {\bf 100}, 210401  (2008).

\bibitem{Sinatra_2009}
Y.,  Li, P. Treutlein, J. Reichel and A. Sinatra,
\newblock Spin squeezing in a bimodal condensate: spatial dynamics and particle losses,
\newblock {\it The European Physical Journal B}, {\bf 68}, 365–381 (2009).

\bibitem{PRA_Poulsen_2002}
U.~V. Poulsen Schulte and K. M\o{}lmer,
\newblock Quantum beam splitter for atoms,
\newblock {\it Phys. Rev. A}, {\bf 65}, 033613 (2002).

\bibitem{PRA_Burchianti_2020}
A. Burchianti, C. D'Errico, L. Marconi, F. Minardi, C. Fort, and M. Modugno,
\newblock Effect of interactions in the interference pattern of Bose-Einstein condensates,
\newblock {\it Phys. Rev. A}, {\bf 102}, 043314 (2020).

\bibitem{PRA_Pezze_2006}
L. Pezzè, A. Smerzi, G.~P. Berman, A.~R. Bishop and L.~A. Collins,
\newblock Nonlinear beam splitter in Bose-Einstein-condensate interferometers,
\newblock {\it Phys. Rev. A}, {\bf 74}, 033610 (2006).

\bibitem{PRA_Roy_2016}
R.~Roy, A.~Green, R.~Bowler, and S.~Gupta,
\newblock Rapid cooling to quantum degeneracy in dynamically shaped atom traps,
\newblock {\it Phys. Rev. A}, {\bf 93}, 043403 (2016).

\bibitem{PRL_Condon_2019}
G. Condon, M. Rabault, B. Barrett, L. Chichet, R. Arguel, H. Eneriz-Imaz, D. Naik, A. Bertoldi, B. Battelier, P. Bouyer, and A. Landragin,
\newblock All-Optical Bose-Einstein Condensates in Microgravity
,
\newblock {\it Phys. Rev. Lett.}, {\bf 123}, 240402  (2019).

\bibitem{Freier2016}
C.~Freier, M.~Hauth, V.~Schkolnik, B.~Leykauf, M.~Schilling, H.~Wziontek, H.-G. Scherneck, J.~M{\"u}ller, and A.~Peters,
\newblock Mobile quantum gravity sensor with unprecedented stability,
\newblock {\it Journal of Physics: Conference Series} {\bf 723}, 012050 (2016).

\bibitem{Karcher2018}
R.~Karcher, A.~Imanaliev, S.~Merlet, and F.~Pereira~Dos Santos,
\newblock Improving the accuracy of atom interferometers with ultracold
  sources,
\newblock {\it New J. of Phys.} {\bf 20}, 113041 (2018).

\bibitem{Battelier2019}
B. Battelier, et al., 
\newblock Exploring the Foundations of the Universe with Space Tests of the Equivalence Principle,
\newblock arXiv:1908.11785.

\bibitem{Tino_2021b}
G. M. Tino, L.~Cacciapuoti, S.~Capozziello, G.~Lambiase, and F.~Sorrentino,
Precision gravity tests and the Einstein equivalence principle,
{\it Progress in Particle and Nuclear Physics} {\bf 112}, 103772 (2020).

\bibitem{Canuel2020}
B.~Canuel, et al., 
\newblock {{ELGAR}}\textemdash a {{European Laboratory}} for {{Gravitation}}
  and {{Atom}}-interferometric {{Research}},
\newblock {\it Classical and Quantum Gravity} {\bf 37}, 225017 (2020).

\end{thebibliography}
\end{document}